\documentclass[transaction,twoside]{IEEEtran}
\usepackage{graphicx,epsfig}
\usepackage{amssymb}
\usepackage{amsmath}
\usepackage{cite}
\usepackage{stfloats}
\usepackage{subfigure}
\usepackage{epstopdf}
\usepackage{mdwlist}
\usepackage[section]{algorithm}
\usepackage[noend]{algpseudocode}

\usepackage[all,2cell]{xy} \UseAllTwocells
\usepackage{booktabs}

\usepackage{etex}
\usepackage{tikz}
\usepackage{xcolor}
\usepackage{enumitem}
\usepackage{nicefrac}

\usepackage{multicol}

\newtheorem{theorem}{Theorem}
\newtheorem{lemma}{Lemma}

\newtheorem{proof}{Proof}

\AtBeginDocument{}

\newcommand{\beq}{\begin{equation}}
\newcommand{\eeq}{\end{equation}}
\newcommand{\beqa}{\begin{eqnarray}}
\newcommand{\eeqa}{\end{eqnarray}}

\begin{document}

\title{
        \hspace{2cm}\\[-0.8cm]
        Power Allocation in Cache-Aided NOMA Systems: Optimization and Deep Reinforcement Learning Approaches
        }

\author{
	Khai Nguyen Doan,~\IEEEmembership{Student Member,~IEEE},
	Mojtaba Vaezi,~\IEEEmembership{Senior Member,~IEEE},
	Wonjae Shin,~\IEEEmembership{Member,~IEEE},
	H. Vincent Poor,~\IEEEmembership{Fellow,~IEEE},
	Hyundong Shin,~\IEEEmembership{Senior Member,~IEEE}
	and
	Tony Q. S. Quek,~\IEEEmembership{Fellow,~IEEE} \\
	\thanks{This work was supported in part by the U.S. National Science Foundation under Grants CCF-0939370 and CCF-1513915, and Basic Science Research Program through the National Research Foundation of Korea (NRF) funded by the Ministry of Science and ICT (NRF-2019R1C1C1006806).}
	\thanks{
		K.~N.~Doan is with the
		Singapore University of Technology and Design, Singapore
		(e-mail: nguyenkhai\textunderscore doan@mymail.sutd.edu.sg). 
	}
	\thanks{
		M. Vaezi is with the
		Department of Electrical and Computer Engineering, Villanova University, Villanova, PA, USA (e-mail: mvaezi@villanova.edu).
	}
	\thanks{
		W. Shin is with the
		Department of Electronics Engineering, Pusan National University, Busan, South Korea and also with Department of Electrical Engineering, Princeton University, Princeton, NJ, USA (e-mail: wjshin@pusan.ac.kr).
	}
	\thanks{
		H. V. Poor is with the
		Department of Electrical Engineering, Princeton University, Princeton, NJ, USA (e-mail: poor@princeton.edu).
	}
	\thanks{
		H.~Shin is with the Department of Electronics and Radio Engineering,
		Kyung Hee University, Yongin-si, Gyeonggi-do, Korea (e-mail: hshin@khu.ac.kr).
	}
	\thanks{
		T. Q. S. Quek is with the Singapore University of Technology and Design, Singapore 487372, and also with the Department of Electronic Engineering, Kyung Hee University, Yongin 17104, South Korea (e-mail: tonyquek@sutd.edu.sg).
	}
}


\maketitle

\vspace{-1.5cm}

\begin{abstract}

This work exploits the advantages of two prominent techniques in future communication networks, namely caching and non-orthogonal multiple access (NOMA). Particularly, a system with Rayleigh fading channels and cache-enabled users is analyzed. It is shown that the caching-NOMA combination provides a new opportunity of cache hit which enhances the cache utility as well as the effectiveness of NOMA. Importantly, this comes without requiring users' collaboration, and thus, avoids many complicated issues such as users' privacy and security, selfishness, etc. In order to optimize users' quality of service and, concurrently, ensure the fairness among users, the probability that all users can decode the desired signals is maximized. In NOMA, a combination of multiple messages are sent to users, and the defined objective is approached by finding an appropriate power allocation for message signals. To address the power allocation problem, two novel methods are proposed. The first one is a divide-and-conquer-based method for which closed-form expressions for the optimal resource allocation policy are derived making this method simple and flexible to the system context. The second one is based on deep reinforcement learning method that allows all users to share the full bandwidth. Finally, simulation results are provided to demonstrate the effectiveness of the proposed methods and to compare their performance.

\end{abstract}

%
%

\section{Introduction}

The dramatic growth in the number and capabilities of mobile devices has triggered a dramatic increase in demand for data over wireless networks \cite{vaezi2018multiple}. This issue is causing a massive load on the backhaul of such networks, especially in systems with densely deployed access points \cite{CAG:08:COMMAG}, and hence, seriously affecting user quality of service (QoS). Proactive caching techniques provide a promising solution to this problem, in which, besides caching at BS \cite{DNSQ:18}, proactively forwarding content to users' devices has been shown to achieve significant benefits \cite{YCTY:16:TWC,MGA_1,DNSQ:Access}. An obvious advantage of this technique is that it offers opportunities for users to retrieve desired content right from their devices, which considerably reduces the number of transmission sessions needed at peak traffic times, thus, saving peak power and bandwidth. 

Non-orthogonal multiple access (NOMA) is another solution for enhancing system capacity and user experiences. NOMA outperforms its counterpart, orthogonal multiple access (OMA), in many contexts by achieving higher power efficiency and lower spectrum usage \cite{MWV:16:MAG,DBLP:journals/corr/WeiYNED16,ZMH:15:CL}. One of the well-known methods to deploy NOMA is based on the power domain in which multiple users' signals are superposed with different power levels while users share a common radio resource of time and frequency \cite{YYATAK:13:VTC,FHJV:16:TC,ZLJPX_4}. Subsequently, successive interference cancellation (SIC) is applied by the receivers to decode the desired information.

The deployment of NOMA has been considered in an attempt to find optimal power allocation policies \cite{DSVPQ:18,DBLP:journals/corr/ZhuWHHYY17}. However there are very few works investigating potential benefits of caching in NOMA context. When enhancing the system performance with the involvement of caching, the situation will be different and more complicated. The reason is that users now are not only affected by the channel conditions, but also the cache placement at the time of generating requests. Because the cached content can be used to eliminate (part of) the interference in the superposed signal. In terms of these techniques combination, \cite{DBLP:journals/corr/abs-1709-06951} jointly considered the advantages of caching and NOMA. This work designed a power allocation method to ensure that the most popular files could be obtained by a predefined number of content servers. In the recent work \cite{YFu18}, the authors narrowed their analysis to a specific case when the user with weaker channel cached information of the user with stronger channel. In \cite{YFu19}, the authors focused on minimizing the power consumption in the system. From another point of view, designing a power allocation policy to maximize the users' QoS as well as to guarantee fairness among users is necessary. Moreover, exploiting users' cached content for interference cancellation can improve users achievable rates and should be paid sufficient attention. However, these points have not been jointly considered in the aforementioned works and many of the previous works \cite{CachingNOMA_2,CachingNOMA_3}. In addition, in NOMA systems without caching, power is typically allocated in inverse order of users' channel conditions \cite{vaezi2019non}. However, the user having the worst channel condition may experience the lowest level of interference thanks to the interference cancellation capability offered by caching technique. Therefore, power allocation schemes in non-caching NOMA systems may not be optimal anymore in this context. Thus, it is useful to find resource allocation policies in which the assumptions on the order of channels and the order of allocated power is relaxed. Inspired by these, we propose two power allocation methods to maximize the users' success probability defined as the probability that all users successfully decode their desired signals in a cache-enabled NOMA system. Furthermore, each user can cache a variety of different content items, resulting in many different situations, and our methods are to work in all of those cases. The main contributions of this work are listed as follows:
\begin{itemize}
	\item We propose a method in which users are paired and the user pairs are separated by orthogonal subchannels to reduce interference. Then, closed-form expressions of power sharing for every user are derived. This method makes use of the channel gain distribution knowledge, users' cached and requested content items at the instant time to maximize the success probability.
	\item Also, from the spectrum efficiency perspective, another solution is proposed that allows all users to share the full bandwidth. In this context, we formulate the problem as a mixed-integer programming problem which can be solved by existed mixed-integer programming algorithms \cite{MI_Prog_6,bookMIP_5,bookMIP_7}.
	\item In order to avoid time-consuming iterative algorithms for solving the formulated optimization problem, a deep-learning-based power allocation method is proposed. Regarding this, we follow a reinforcement learning approach that improves the performance by observing the accuracy of each applied power allocation pattern. We then propose a dual deep neural network model to deal with the nosiness/randomness in the collected training data.
	
\end{itemize}  

The remainder of this work is organized as follows. Section~\ref{Sec:SystemModel} describes the system model under consideration. Section~\ref{Sec:Method1} presents our first power allocation method. The second power allocation method is described in Section~\ref{Sec:Method2} where we first formulate the problem as a solvable mixed-integer programming and then cast it into another form to fit into our designed learning scheme. Subsequently, our simulation results are shown in Section~\ref{Sec:Results}. Finally, Section~\ref{Sec:Conclusion} concludes our work.

\section{System Model} \label{Sec:SystemModel}

\begin{figure}[ht]
	\centering
	\includegraphics[scale = 0.19]{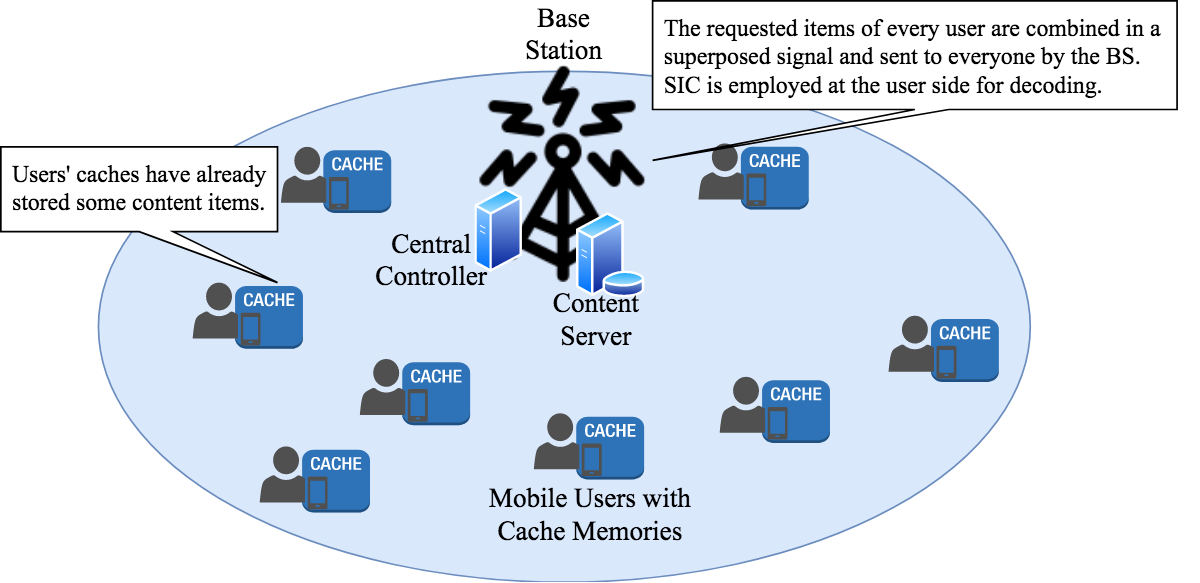}
	\caption{The system model under consideration consisting of a BS with a content server serving a set of $K$ users. Each user's device is implemented with a cache storage. At the beginning, each user has already cached some content items.}
	\label{fig:Sys_Mod}
\end{figure}

We consider a system consisting of $K$ users served by a BS having a content server. Each user's device has a cache with finite capacity and we assume that users cache an entire file rather than a partition of it. With the cache, users typically fetch and store a set of files during an off-peak time called \textit{caching phase}. In this work, we assume that the caching phase has already taken place and consider the next stage called \textit{requesting phase}. In this phase, each user  requests for a file in the server. In addition, since files are sent to users for caching by the BS, it has information about files placed in users' caches. 

In a downlink NOMA framework in which a superposed signal is transmitted by the BS to users, the SIC is employed at each user to decode the desired signal. Let $P_{\max}$ be the total transmission power and $\alpha_i$ be the portion allocated to the signal of user $i$ where $0 \le \alpha_i \le 1, \forall i=1,\ldots,K$. With SIC, user $i$ will decode a sequence of signals from the one with strongest to weakest power until the $i$th signal is decoded. This is because, firstly, strong signals are always easier to be decoded than weak signals. Secondly, decoding and removing stronger signals from the interference will increase the SINR associated with weaker signals making them easier to be decoded. The BS can acknowledge users about decoding order by adding an additional field to the sending information. The model for the signal received at user $i$ is
\begin{align}
y_i = \frac{h_i}{d_i^{\gamma/2}}\sum_jx_j\sqrt{\alpha_jP_{\max}} + n_i
\end{align}
where $x_j$ is the signal corresponding to file $j$; $h_i$ and $d_i$ are the channel coefficient and the distance between user $i$ and the BS, respectively; $\gamma$ is the pathloss exponent; and $n_i$ is the Gaussian noise with mean $0$ and variance $\sigma^2$. Denote $\rho=\frac{P_{\max}}{\sigma^2}$ the signal-to-noise ratio (SNR), and $\beta_i = d_i^\gamma\sigma^2$. $h_i$ follows Rayleigh distribution with parameter $\sigma_i$, and thus $\left|h_i\right|^2$ follows exponential distribution with parameter $\lambda_i = 2/\sigma_i^2$. Note that only the distribution of channels is known by the BS. Let $\mathcal{C}$ be the index set of files cached by user $i$, then, user $i$ can decode the signal $j$ (signal associated with user $j$) when
\begin{align}
\label{general_success_cond}
\frac{\left|h_i \right|^2\alpha_jP_{\max}}{\left|h_i \right|^2\sum_{k \notin \mathcal{C},\alpha_k \le \alpha_j}\alpha_kP_{\max} + \beta_i} \ge \epsilon_j
\end{align}
where $\epsilon_j$ is the minimum SINR required to decode file $j$. The meaning of the interference term in (\ref{general_success_cond}) is that only signals bearing information which has not been cached will constitute the interference; otherwise, those signals will be removed. Thus, we have the condition $k \notin \mathcal{C}$ in the sum. Besides, with SIC, signals with stronger power will be decoded first and removed from the superposed signal. Therefore, when a user decode signal $j$, only weaker signals constitute the interference, and thus we have the condition $\alpha_k \le \alpha_j$ in the range of the sum.

In summary, with SIC users may need to decode a sequence of signals before obtaining their desired ones. In this case, the failure event is said to occur if users fail to decode one of those signals. However, if some content items have already been cached, users can remove the corresponding signals from the interference without decoding, which increases the success probability. Therefore, we aim to design a power allocation policy that exploits both information about channel conditions and cache placement of users to maximize the success probability. Regarding this, we present two power allocation methods in the following two sections. The first method in Section \ref{Sec:Method1} makes use of orthogonal channels to reduce the interference among users, while the second method in Section \ref{Sec:Method2} allows all users to share the whole bandwidth.

By abusing the notations we will use $\epsilon_j$ to denote the minimum SINR level required to decode file $j$ and $\sigma^2$ for the noise power. Note that although the same notation is used for the SINR threshold, this quantity in Section \ref{Sec:Method1} and \ref{Sec:Method2} are not the same. This is because the first method requires subchannel allocation, while the second method does not. If $\epsilon$ is the SINR threshold to decode a file when a user can use the whole bandwidth, and $\epsilon_W$ is that when the user can only use $1/W$ of the bandwidth, then, their relation is as follows
\begin{align}
\label{epsilon_relation}
\epsilon_W = \exp\left(W\log\left(1+\epsilon\right)\right) - 1.
\end{align} 
In addition, if $\sigma^2$ is the power of the Gaussian noise when users use the whole bandwidth, then $\sigma^2/W$ is that when they only use $1/W$ of the bandwidth. In Fig.~\ref{fig:Sys_Mod}, the central controller is a computational unit placed at the BS. This unit gathers and processes information to allocate appropriate power to users' signals.

\section{Method 1: Divide-and-Conquer-Based Optimization Scheme} \label{Sec:Method1}
\begin{figure*}[ht]
	\centering
	\includegraphics[scale = 0.39]{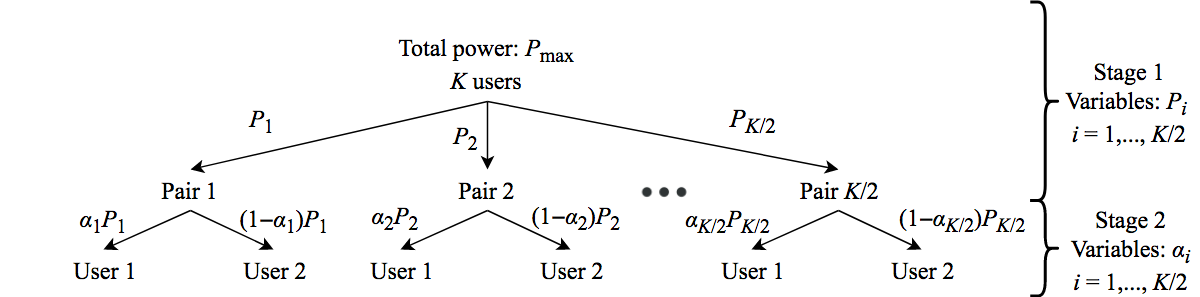}
	\caption{The two-stage power allocation process in which users are paired and user pairs are assigned orthogonal frequencies.}
	\label{fig:twostage_poweralloc}
\end{figure*}

In this method, user pairs are interference-isolated with orthogonal frequencies. The entire bandwidth is evenly divided among all pairs. The time-frequency resource allocation in a multi-carrier setting is an interesting topic; however, it is beyond the scope of this work and is a topic of further study. NOMA is then separately applied to each user pair. In this regime, the power allocation consists of two stages as in Fig.~\ref{fig:twostage_poweralloc}. The first stage is to share the total transmission power $P_{\max}$ to user pairs, so-called inter-pair power allocation stage. We denote $P_i$ to be the power allocated to the $i$th pair where $\sum_{i=1}^{K/2} P_i = P_{\max}$. Then, the second stage is to allocate portions of a given power amount to users in each pair, i.e., the first user in pair $i$ will be allocated with a portion $\alpha_i$ of $P_i$, while the second one is given $1-\alpha_i$. This stage is called intra-pair power allocation stage. \footnote{The divide-and-conquer-based optimization method is also presented in \cite{DSVPQ:18} - a conference version of this work.}

\begin{lemma}
	The whole process can be optimized by optimizing the two stages separately.
\end{lemma}

In order to prove the above lemma, we first describe the problem formulation as follows
\begin{align}
\label{obj_full}
& \max~ \prod_{i=1}^{K/2} \mathcal{G}_i\left(\alpha_i, P_i\right) \\
& \text{s.t. } \sum_{i=1}^{K/2} P_i = P_{\max} \label{coupling}\\
& ~~~~~ P_i \ge 0, \forall i = 1, 2, \ldots, K/2 \label{cond2} \\
& ~~~~~ 0 \le \alpha_i \le 1, \forall i = 1, 2, \ldots, K/2 . \label{obj_full_lastcon}
\end{align}
where $\mathcal{G}_i\left(\alpha_i, P_i\right)$ is the success probability of the $i$th user pair. Through subsection \ref{twousers} (specifically, expressions (\ref{ez1}), (\ref{ez2}), (\ref{spC11}), (\ref{spC12}), (\ref{spC21}), (\ref{spC22}), (\ref{spC3}), (\ref{spC41}) and (\ref{spC42})) it will be shown that, in all cases, the success probability of each user pair takes the following form
\begin{align}
\mathcal{G}_i\left(\alpha_i, P_i\right) = \exp\left(-\frac{\Psi_i\left(\alpha_i\right)}{P_i}\right)
\end{align}
where $\Psi_i\left(\alpha_i\right)$ is a function of $\alpha_i$ and does not depend on $P_i$. In the above formulation, \eqref{obj_full} is the probability that all users are success. In the subsection below, we will alternatively analyze all possible cases in which the expressions of user pair's success probability are different in different cases. However, the problem formulations in all cases share a common form which is \eqref{obj_full}-\eqref{obj_full_lastcon}. Besides that, (\ref{coupling}) is the only coupling constraint, meanwhile $\alpha_i$ and $\alpha_j, \forall i \ne j$ are independent due to orthogonal subchannel assumption. Therefore, the intra-pair (regarding variables $\alpha_i$) and inter-pair (regarding variables $P_i$) power allocation stages can be optimized separately by minimizing $\Psi_i\left(\alpha_i\right) \forall i$ separately, then solving the above problem with $P_i~\forall i$ as variables and with the optimal values of $\alpha_i$ plugged in. In the following, the steps of optimizing $\alpha_i$ are given in subsection \ref{twousers}, while the form of optimal value for $P_i$ is presented in subsection \ref{multipairs}.

\subsection{Intra-Pair Power Allocation Stage} \label{twousers}
In this subsection, we work with the second stage of resource assignment and consider a specific user pair. Hence, the user pair indexes in notations will be dropped out for simplicity. Let us denote $P$ to be the total power allocated to this pair and $0 \le \alpha \le 1$ be the portion of $P$ allocated to the first user, and thus, the portion of the second user is $1-\alpha$. Note that the value of $P$ will be optimized in the first stage.

Without loss of generality, we assume that user 1 requests for file $f_1$ and user 2 requests for file $f_2$. In SIC, for each specific user, the signal with stronger power will be decoded first. Therefore, the main idea is to optimize the success probability for both of the cases, $0 \le \alpha < 0.5$ and $0.5 \le \alpha \le 1$, then, the better one will be chosen.

There are some situations in which the optimal power allocation is trivial, whereas some situations require solving optimization problems to find the optimal policy. Therefore, we will first clear out the trivial cases before addressing the rest. Those trivial cases and the corresponding optimal power allocation policies are as follows
\begin{enumerate}
    \item When both users can be served locally with their caches, no over-the-air transmission is required.
    \item When only one user finds his request in his own cache. We assume that is user 2, then all the power $P$ will be allocated for user 1. User 1 can decode his file successfully when
        \begin{align}
        \frac{\left|h_1\right|^2}{\beta_1} \ge \epsilon_1
        \end{align}
        Because $\left|h_i\right|^2 \sim Exp\left(\lambda_i\right)$, the success probability is
        \begin{align}
        \label{ez1}
        \Pr\left\{\left|h_1\right|^2 \ge \epsilon_1\beta_1\right\} = \exp\left(- \frac{\lambda_1\epsilon_1\beta_1}{P} \right).
        \end{align}
    \item When both users request for the same file, but neither of them have cached it. Then, a single signal representing that file is sent with power $P$ to both users. Both users can successfully decode that file when
        \begin{align}
        & \frac{\left|h_1\right|^2}{\beta_1} \ge \epsilon_{1,2} \\
        & \frac{\left|h_2\right|^2}{\beta_2} \ge \epsilon_{1,2}
        \end{align}
        where $\epsilon_{1,2}$ denotes the SINR threshold of the file requested by both users. The success probability is given by
        \begin{align}
        \label{ez2}
        \begin{split}
        \Pr\left\{\left|h_1\right|^2 \ge \epsilon_{1,2}\beta_1, \left|h_2\right|^2 \ge \epsilon_{1,2}\beta_2 \right\}\\
        = \exp\left(-\frac{\epsilon_{1,2}}{P}\left(\lambda_1\beta_1 + \lambda_2\beta_2 \right) \right).
        \end{split}
        \end{align}
\end{enumerate}

To this end, it remains to consider four more-complicated cases. Without loss of generality, we assume that
\begin{align}
\label{zeta_def}
\zeta = \frac{\lambda_1\epsilon_1\beta_1}{\lambda_2\epsilon_2\beta_2} \ge 1.
\end{align}
The remaining cases are listed as follows:
\begin{enumerate}
    \item \textit{C1}: User 1 has cached $f_2$ and user 2 has had a cache miss.
    \item \textit{C2}: User 1 has had a cache miss and user 2 has cached $f_1$.
    \item \textit{C3}: User 1 has cached $f_2$ and user 2 has cached $f_1$.
    \item \textit{C4}: Both users have had cache misses.
\end{enumerate}
where the term ``cache miss" implies that users have not cached any file in the coming signal. Hereafter, we will analyze and derive the optimal power allocation for each case. 

In case $C1$, user 1 is capable of eliminating the interference from the superposed signal by exploiting the cached $f_2$. Thus, user 1 can decode the desired file when
\begin{align}
\label{conC1_1}
\frac{\left|h_1\right|^2\alpha P}{\beta_1} \ge \epsilon_1.
\end{align}
For the second user, if $0.5 \le \alpha \le 1$, user 2 has to decode $f_1$ with signal of $f_2$ as interference, remove $f_1$ from the superposed signal and then decode $f_2$. If $0 \le \alpha \le 0.5$, he can decode $f_2$ directly with $f_1$ as interference. These points can be expressed as follows
\begin{itemize}
    \item If $0.5 \le \alpha \le 1$
    \begin{align}
    & \frac{\left|h_2\right|^2\alpha P}{\left|h_2\right|^2\left(1-\alpha\right)P + \beta_2} \ge \epsilon_1 \label{conC1_2}\\
    & \frac{\left|h_2\right|^2\left(1-\alpha\right)P}{\beta_2} \ge \epsilon_2. \label{conC1_3}
    \end{align}
    With simple manipulating steps and the fact that $\left|h_i\right|^2 \sim Exp\left(\lambda_i\right)$, the success probability can be expressed as
    \begin{equation}
    \label{spC11}
    p^{C_1}_1 = \exp\left(-\frac{1}{P} \times \tilde{p}^{C_1}_1\right),
    \end{equation}
	where
	\begin{align}
	\tilde{p}^{C_1}_1 = \frac{\lambda_1\epsilon_1\beta_1}{\alpha} + \max\left(\frac{\lambda_2\epsilon_2\beta_2}{1-\alpha}, \frac{\lambda_2\epsilon_1\beta_2}{\left(1+\epsilon_1\right)\alpha-\epsilon_1}\right).
	\end{align}
    \item If $0 \le \alpha \le 0.5$
    \begin{align}
    \frac{\left|h_2\right|^2\left(1-\alpha\right)P}{\left|h_2\right|^2\alpha P + \beta_2} \ge \epsilon_2 \label{conC1_4}
    \end{align}
    which gives the following success probability
    \begin{align}
    \label{spC12}
    p^{C_1}_2 = \exp\left(-\frac{1}{P}\left(\frac{\lambda_1\epsilon_1\beta_1}{\alpha} + \frac{\lambda_2\epsilon_2\beta_2}{1-\left(1+\epsilon_2\right)\alpha}\right)\right).
    \end{align}
    
\end{itemize}
In this case, the optimal power allocation is given as in the following theorem.
\begin{theorem}
\label{theorem1}
When user 1 has cached $f_2$ and user 2 has had a cache miss, given $\zeta \ge 1$, the optimal power allocation is
\begin{align}
\alpha^* =
\begin{cases}
z_1^{\textit{C1}} \text{, if } g_1^{\textit{C1}}\left(z_1^{\textit{C1}}\right) \le g_2^{\textit{C1}}\left(z_2^{\textit{C1}} \right) \\
z_2^{\textit{C1}} \text{, otherwise}
\end{cases}
\end{align}
where $g_1^{\textit{C1}}$, $g_2^{\textit{C1}}$, $z_1^{\textit{C1}}$, and $z_2^{\textit{C1}}$ are defined as follows
\begin{align}
& g_1^{\textit{C1}}\left(z\right) = \frac{\lambda_1\epsilon_1\beta_1}{z} + \frac{\lambda_2\epsilon_2\beta_2}{1-z} \label{for_appendix_1} \\
& g_2^{\textit{C1}}\left(z\right) = \frac{\lambda_1\epsilon_1\beta_1}{z} + \frac{\lambda_2\epsilon_2\beta_2}{1 - \left(1 + \epsilon_2 \right)z} \label{for_appendix_2} \\
& z_1^{\textit{C1}} = \max\left(1 - \frac{1}{\sqrt{\zeta}+1}, 1 - \frac{1}{1+\epsilon_1 + \frac{\epsilon_1}{\epsilon_2}} \right) \\
& z_2^{\textit{C1}} = \min\left(\frac{1}{1+\epsilon_2}\left(1 - \frac{1}{\sqrt{\zeta\left(1+\epsilon_2\right)} + 1}\right), 0.5 \right).
\end{align}
\end{theorem}

\textit{Proof: Please see an appendix in Section \ref{Sec:Appendix}.}

Note that the case \textit{C2} is not equivalent to \textit{C1}, due to (\ref{zeta_def}). In \textit{C2}, user 2 removes the interference by using the cached content, thus, can decode the desired signal when
\begin{align}
\label{conC2_1}
\frac{\left|h_2\right|^2\left(1-\alpha\right)P}{\beta_2} \ge \epsilon_2.
\end{align}

Similar to the previous case, the success conditions for user 1 are
\begin{itemize}
    \item If $0.5 \le \alpha \le 1$
    \begin{align}
    \frac{\left|h_1\right|^2\alpha P}{\left|h_1\right|^2\left(1-\alpha\right)P + \beta_1} \ge \epsilon_1 \label{conC2_2}.
    \end{align}
    Similarly to the manipulation in the previous case, we have success probability as
    \begin{align}
    \label{spC21}
    p^{C_2}_1 = \exp\left(-\frac{1}{P}\left(\frac{\lambda_1\epsilon_1\beta_1}{\left(1+\epsilon_1\right)\alpha - \epsilon_1} + \frac{\lambda_2\epsilon_2\beta_2}{1-\alpha} \right)\right).
    \end{align}
    
    \item If $0 \le \alpha \le 0.5$
    \begin{align}
    & \frac{\left|h_1\right|^2\left(1-\alpha\right)P}{\left|h_1\right|^2\alpha P + \beta_1} \ge \epsilon_2 \label{conC2_3} \\
    & \frac{\left|h_1\right|^2\alpha P}{\beta_1} \ge \epsilon_1. \label{conC2_4}
    \end{align}
    The corresponding success probability is
    \begin{align}
    \label{spC22}
    p^{C_2}_2 = \exp\left(-\frac{1}{P} \times \tilde{p}^{C_2}_2\right),
    \end{align}
    where
    \begin{align}
    \tilde{p}^{C_2}_2 = \max\left(\frac{\lambda_1\epsilon_2\beta_1}{1-\left(1+\epsilon_2\right)\alpha}, \frac{\lambda_1\epsilon_1\beta_1}{\alpha}\right) + \frac{\lambda_2\epsilon_2\beta_2}{1-\alpha}.
    \end{align}
\end{itemize}
Then, the optimal power allocation for \textit{C2} is presented in Theorem \ref{theorem2}.
\begin{theorem}
\label{theorem2}
When user 1 has had a cache miss and user 2 has cached $f_1$, given $\zeta \ge 1$, the optimal power allocation is as follows
\begin{align}
\alpha^* =
\begin{cases}
z_1^{\textit{C2}} \text{, if }  g_1^{\textit{C2}}\left(z_1^{\textit{C2}}\right) \le g_2^{\textit{C2}}\left(z_2^{\textit{C2}}\right) \\
z_2^{\textit{C2}} \text{, otherwise}
\end{cases}
\end{align}
where $g_1^{\textit{C2}}$, $g_2^{\textit{C2}}$, $z_1^{\textit{C2}}$ and $z_2^{\textit{C2}}$ are defined as follows
\begin{align}
& g_1^{\textit{C2}}\left(z\right) = \frac{\lambda_1\epsilon_1\beta_1}{\left(1+\epsilon_1 \right)z - \epsilon_1} + \frac{\lambda_2\epsilon_2\beta_2}{1-z} \\
& g_2^{\textit{C2}}\left(z\right) = \frac{\lambda_1\epsilon_1\beta_1}{z} + \frac{\lambda_2\epsilon_2\beta_2}{1-z} \\
& z_1^{\textit{C2}} = 1 - \frac{1}{\sqrt{\zeta\left(1+\epsilon_1\right)} + 1+\epsilon_1 } \\
& z_2^{\textit{C2}} = \min\left(\frac{1}{1+\epsilon_2 + \frac{\epsilon_2}{\epsilon_1}}, 0.5\right).
\end{align}
\end{theorem}

Next, for the case \textit{C3}, both users can use their cached content items to remove the interference from the superposed signal, therefore, the conditions for them to successfully decode the desired signal does not depend on where $\alpha$ is relative to $0.5$.
\begin{align}
& \frac{\left|h_1\right|^2\alpha P}{\beta_1} \ge \epsilon_1 \label{conC3_1}\\
& \frac{\left|h_2\right|^2\left(1-\alpha\right)P}{\beta_2} \ge \epsilon_2. \label{conC3_2}
\end{align}
The success probability of this case can be derived as follows
\begin{align}
\label{spC3}
p^{C_3} = \exp\left(-\frac{1}{P}\left(\frac{\lambda_1\epsilon_1\beta_1}{\alpha} + \frac{\lambda_2\epsilon_2\beta_2}{1-\alpha}\right)\right).
\end{align}
Then, the optimal policy for this case is simple and summarized in the next theorem.
\begin{theorem}
\label{theorem3}
When user 1 has cached $f_1$ and user 2 has cached $f_2$, given $\zeta \ge 1$, the optimal power allocation is as follows
\begin{align}
\alpha^* = 1 - \frac{1}{\sqrt{\zeta} + 1}.
\end{align}
\end{theorem}

Finally, case \textit{C4}, both users experience interference, and user with lower power need to decode other users' file, remove it from the superposed signal before decoding his own file. Thus, the success conditions are
\begin{itemize}
\item If $0.5 \le \alpha \le 1$
\begin{align}
& \frac{\left|h_1\right|^2\alpha P}{\left|h_1\right|^2\left(1-\alpha\right)P + \beta_1} \ge \epsilon_1 \label{conC4_1}\\
& \frac{\left|h_2\right|^2\alpha P}{\left|h_2\right|^2\left(1-\alpha\right)P + \beta_2} \ge \epsilon_1 \label{conC4_2}\\
& \frac{\left|h_2\right|^2\left(1-\alpha\right)P}{\beta_2} \ge \epsilon_2. \label{conC4_3}
\end{align}
with similar manipulation steps as case $C1$, the success probability is derived as
\begin{align}
\label{spC41}
p^{C4}_1 = \exp\left(-\frac{1}{P}\times \tilde{p}^{C4}_1\right),
\end{align}
where
\begin{align}
\begin{split}
& \tilde{p}^{C4}_1 = \\ 
& \frac{\lambda_1\epsilon_1\beta_1}{\left(1+\epsilon_1\right)\alpha-\epsilon_1} + \max\left(\frac{\lambda_2\epsilon_1\beta_2}{\left(1+\epsilon_1\right)\alpha-\epsilon_1}, \frac{\lambda_2\epsilon_2\beta_2}{1-\alpha}\right).
\end{split}
\end{align}
\item If $0 \le \alpha \le 0.5$
\begin{align}
& \frac{\left|h_1\right|^2\left(1-\alpha\right)P}{\left|h_1\right|^2\alpha P + \beta_1} \ge \epsilon_2 \label{conC4_4}\\
& \frac{\left|h_1\right|^2\alpha P}{\beta_1} \ge \epsilon_1 \label{conC4_5} \\
& \frac{\left|h_2\right|^2\left(1-\alpha\right)P}{\left|h_2\right|^2\alpha P + \beta_2} \ge \epsilon_2. \label{conC4_6}
\end{align}
The corresponding success probability is
\begin{align}
\label{spC42}
p^{C4}_2 = \exp\left(-\frac{1}{P} \times \tilde{p}^{C4}_2\right),
\end{align}
where
\begin{align}
\begin{split}
&\tilde{p}^{C4}_2 = \\ &\max\left(\frac{\lambda_1\epsilon_2\beta_1}{1-\left(1+\epsilon_2\right)\alpha}, \frac{\lambda_1\epsilon_1\beta_1}{\alpha}\right) + \frac{\lambda_2\epsilon_2\beta_2}{1-\left(1+\epsilon_2\right)\alpha}.
\end{split}
\end{align}
\end{itemize}

Theorem \ref{theorem4} presents the optimal policy for this case.
\begin{theorem}
\label{theorem4}
Given $\zeta \ge 1$, when both users have cache misses, the optimal power allocation policy is as follows
\begin{align}
\alpha^* =
\begin{cases}
z_1^{\textit{C4}} \text{, if } g_1^{\textit{C4}}\left(z_1^{\textit{C4}}\right) \le g_2^{\textit{C4}}\left(z_2^{\textit{C4}} \right) \\
z_2^{\textit{C4}} \text{, otherwise}
\end{cases}
\end{align}
where $g_1^{\textit{C4}}$, $g_2^{\textit{C4}}$, $z_1^{\textit{C4}}$ and $z_2^{\textit{C4}}$ are defined as follows
\begin{align}
& g_1^{\textit{C4}}\left(z\right) = \frac{\lambda_1\epsilon_1\beta_1}{\left(1+\epsilon_1 \right)z - \epsilon_1} + \frac{\lambda_2\epsilon_2\beta_2}{1-z} \\
& g_2^{\textit{C4}}\left(z\right) = \frac{\lambda_1\epsilon_1\beta_1}{z} + \frac{\lambda_2\epsilon_2\beta_2}{1 - \left(1+\epsilon_2\right)z} \\
& z_1^{\textit{C4}} = 1 - \min\left(\frac{1}{\sqrt{1+\epsilon_1}\left(\sqrt{\zeta} + \sqrt{1+\epsilon_1} \right)}, \frac{1}{1 + \epsilon_1 + \frac{\epsilon_1}{\epsilon_2}} \right) \\
& z_2^{\textit{C4}} = \min\left(\frac{1 - \frac{1}{\sqrt{\zeta\left(1 + \epsilon_2 \right)} + 1} }{1+\epsilon_2}, \frac{1}{1 + \epsilon_2 + \frac{\epsilon_2}{\epsilon_1}}, 0.5 \right).
\end{align}
\end{theorem}

Note that the expression of $z_2^{\textit{C4}}$ is a minimum function with three arguments. Due to space limitation, the proofs of Theorem \ref{theorem2}, \ref{theorem3} and \ref{theorem4} are omitted, however, their results can be derived in a very similar way as that of Theorem \ref{theorem1}.

In summary, we have presented, in this subsection, the optimal power allocation for each user in a pair where user pairs are given orthogonal subchannels. The system state information constituted by users' requests and their cached content items can be gathered by the BS at the time requests are generated. Then, the current state can be defined to be one of the analyzed cases, and the corresponding power allocation is applied. In this subsection, we assume that the total power given to the considered user pair is fixed to be $P$, and the optimal portion shared to each user is derived. In the next subsection, we will discuss how $P$ is defined for each pair of user. Before closing this subsection, we will point out some important observations.

\subsection{Inter-Pair Power Allocation Stage} \label{multipairs}

In this subsection, we address the power allocation for user pairs, i.e., deriving $P_i$ for all pair $i$ such that the success probability is maximized. As pointed out at the beginning of this subsection, when optimizing variables $P_i$, $\Psi_i \forall i$ are fixed at their optimal values which are obtained by substituting $\alpha_i$ by the results given in Theorem \ref{theorem1}-\ref{theorem4}. Let us denote the optimal value of $\Psi_i\left(\alpha_i\right)$ (when $\alpha_i$ is optimized) as $\Psi^*_i$.

To this end, the objective function in (\ref{obj_full}) can be written as
\begin{align}
\mathcal{G}\left(\mathbf{P}\right) = \exp\left(-\sum_{i=1}^{K/2}\frac{\Psi^*_i}{P_i}\right)
\end{align}
where $\mathbf{P}=\left[P_1, P_2, \ldots, P_{K/2}\right]$ and (\ref{obj_full}) can be replaced by
\begin{align}
\label{obj_alter}
\underset{\mathbf{P}}{\min}~ \sum_{i=1}^{K/2}\frac{\Psi^*_i}{P_i}.
\end{align}

Solving (\ref{obj_alter}) with the set of constraints (\ref{coupling}) and (\ref{cond2}) by applying KKT conditions gives us the following closed-form solution
\begin{align}
\label{P_for_pairs}
P_i = \frac{\sqrt{\Psi^*_i}}{\sum_{j=1}^{K/2}\sqrt{\Psi^*_j}}P_{\max}, \forall i=1,\ldots,K/2.
\end{align}

Note that each user pair is assigned $2/K$ of the available bandwidth, therefore, as mentioned at the end of Section \ref{Sec:SystemModel}, the SINR thresholds required to decode files will increase as the number of user pairs increases, however, the noise power $\sigma^2$ will decrease for each pair. This is because $\sigma^2$ is inversely proportional to the number of partitioned subchannels. 

In summary, the method proposed in this section helps create two separable power allocation stages. The closed-form solutions for both stages are provided in Theorem \ref{theorem1}-\ref{theorem4}, three trivial cases and (\ref{P_for_pairs}). These results together with the simple design of this method allow it to be flexibly applicable in various system contexts.

\section{Method 2: Deep-Reinforcement-Learning-Based Scheme} \label{Sec:Method2}

Our presented Method 1 simplifies the problem with a user pairing technique that allows us to obtain closed-form expressions. In this section, from the bandwidth efficiency perspective, we propose another method based on machine learning, which not only responds quickly upon users' requests but also allows all users to share the entire bandwidth. The application and performance of both methods will be compared and summarized in Sections \ref{Sec:Results} and \ref{Sec:Conclusion}. However, before discussing the technical details, we will formulate the power allocation problem as an optimization problem. This is, first, to provide a better mathematical view of our considered problem. Second, we want to show that applying iterative optimization algorithms is not suitable for this context since users require a short-delay response from the BS. These are the main motivations for proposing a learning-based approach.

\subsection{Problem Formulation}
For the ease of notation, let us number the users from 1 to $K$, and we also call $f_1, f_2, \ldots, f_K$ the files requested by user 1, 2, $\ldots, K$, respectively. Let $C_{ij} = 1$ if the $i$th user has cached $f_j$, and $C_{ij}=0$, otherwise, $\forall i,j = 1,\ldots, K$. Similar to the previous section, $C_{ij}, \forall i,j$ are known prior to the power allocation process. Recall that the success probability expression is not defined if the order of $\alpha_1, \alpha_2, \ldots, \alpha_K$ is not defined, where $\alpha_i$ is the portion of the total power $P_{\max}$ allocated to the $i$th user. Therefore, we use another set of variables $\psi_{ij}$ which, by taking value 1, implying that $\alpha_i\ge \alpha_j$ and by taking value 0, implying that $\alpha_i < \alpha_j$.

The condition for user $i$ to successfully decodes file $f_j$ is as follows
\begin{align}
\frac{\left|h_i\right|^2\alpha_j}{\left|h_i\right|^2\sum_{k\ne j}C_{ik}\psi_{jk}\alpha_k + \beta_i} - \left(1-C_{ij}\right)\epsilon_j \ge 0
\end{align}
which can be rewritten as
\begin{align}
\label{suc_cond1}
\begin{cases}
\left|h_i\right|^2 \ge \frac{\left(1-C_{ij}\right)\epsilon_j\beta_i}{\alpha_j - \epsilon_j\sum_{k\ne j}C_{ik}\psi_{jk}\alpha_k} \\
\alpha_j - \epsilon_j\sum_{k\ne j}C_{ik}\psi_{jk}\alpha_k > 0
\end{cases}
.
\end{align}

In other words, if (\ref{suc_cond1}) are not satisfied, user $i$ cannot decode file $f_j$. Maximizing the success probability is maximizing the following function
\begin{align}
\label{obj_temp}
\exp\left(-\sum_{i=1}^K \lambda_i \sum_{j=1}^K \frac{\left(1-C_{ij}\right)\left(1-\psi_{ij}\right)\epsilon_j\beta_i}{\alpha_j  - \epsilon_j\sum_{k\ne j}C_{ik}\psi_{jk}\alpha_k}\right)
\end{align}
where the term $\left(1-\psi_{ij}\right)$ is added to the objective function to imply that the $i$-th user only need to obtain file $f_j$ if $\alpha_i < \alpha_j$ or equivalently, $\psi_{ij}=0$. The binary variables $\psi_{ij}$ are to define the order of $\alpha_i, \forall i,j$. To this end, the problem formulation to maximize the success probability can be expressed as
\begin{align}
& \underset{\psi,\alpha}{\min}~ \sum_{i=1}^K \lambda_i \sum_{j=1}^K \frac{\left(1-C_{ij}\right)\left(1-\psi_{ij}\right)\epsilon_j\beta_i}{\alpha_j  - \epsilon_j\sum_{k\ne j}C_{ik}\psi_{jk}\alpha_k} \label{formu_sec2_1}\\
& \text{s.t. }\psi_{ij} + \psi_{ji} = 1, \forall i,j = 1, \ldots, K \label{sec2_con1_formu1} \\
& ~~~~ \psi_{ij} \in \{0,1\}, \forall i,j = 1, \ldots, K \label{sec2_con2_formu1}\\
& ~~~~ \sum_{i=1}^K\alpha_i = 1 \label{sec2_con3_formu1}\\
& ~~~~ -1 \le \alpha_i - \alpha_j - \psi_{ij} \le 0 , \forall i, j = 1,\ldots,K \label{sec2_con4_formu1}\\
& ~~~~ \alpha_j - \epsilon_j\sum_{k\ne j}C_{ik}\psi_{jk}\alpha_k \ge \xi, \forall i,j = 1,\ldots,K \label{con5_formu1}\\
& ~~~~ 0 \le \alpha_i \le 1, \forall i = 1, \ldots, K. \label{sec2_con6_formu1}
\end{align}
where maximizing (\ref{obj_temp}) is equivalent to (\ref{formu_sec2_1}). The constraint set (\ref{sec2_con1_formu1}) is to guarantee the consistency, i.e., as $\alpha_i < \alpha_j$ or $\psi_{ij}=1$, then we cannot have $\alpha_j < \alpha_i$ meaning that $\psi_{ji}$ must be 0. Constraint sets (\ref{sec2_con2_formu1}), (\ref{sec2_con3_formu1}) and (\ref{sec2_con6_formu1}) are due to the definition of our variables. The constraint set (\ref{sec2_con4_formu1}) presents the relationship between $\psi_{ij}$ and $\alpha_i, \forall i,j$, which is $0 \le \alpha_i - \alpha_j \le 1$ if $\psi_{ij}=1$ and $-1 \le \alpha_i - \alpha_j \le 0$ if $\psi_{ij}=0$. Finally, the constraint set (\ref{con5_formu1}) is from the second inequality of (\ref{suc_cond1}). We have replaced 0 on the right-hand side by a small positive number $\xi$ to convert ``$>$" to ''$\ge$" making the problem easier to be solved by optimization techniques. The problem can be solved by applying searching methods or mixed-integer programming algorithms \cite{MI_Prog_6,bookMIP_5,bookMIP_7}.

Unlike the previous section in which we can allocate orthogonal frequencies to each user pair, and since the number of users occupying a frequency is limited to two, we can derive a closed-form expression for the optimal policy. In this part, users share the same frequency and in addition, the order of $\alpha_i,\forall i$ is not defined yet, hence, we have to solve a mixed integer programming problem with a non-convex objective function as in (\ref{formu_sec2_1})-(\ref{sec2_con6_formu1}). Therefore, optimization algorithms may require several iterations, which is time consuming, thus increasing the delay. In other words, by applying optimization techniques, users always have to wait for the central controller to find a suitable power allocation policy via iterative numerical algorithms.

Being motivated by the above, we proposed a learning mechanism allowing our system to improve its performance over time. This approach, after the training stage, will be able to incur no delay in term of resource allocating process.

\subsection{Deep-Reinforcement-Learning-Based Scheme}

\subsubsection{Problem Re-formulation}

Before introducing a new form of problem formulation such that a learning method can be applied, we would like to mention the concept of \textit{users' advantage} which consists of two sides, the user advantage in terms of channel conditions and cached content items. In the case without caching, before power is allocated to users, user advantage is defined only by channel conditions, i.e., a user with better channel condition is considered to have a higher advantage. Then, we allocate higher power for users having less advantage. In our problem, another type of advantage coming from users' cached content items is that users caching more files in the superposed signal have more advantage in terms of interference cancellation. Therefore, in order to allocate appropriate power, the user advantages need to be jointly exploited from both of these sides.

Following a reinforcement learning method, our central controller (at the BS) can be treated as an acting agent and the environment, here, contains all other components such as channels, SINR thresholds of files, users' requests, cached files, etc. In order to capture the aspect of users' advantages, a state is defined as follows. Let $K^2$-dimensional vector $\mathbf{S}$ represent a state. For the ease of understanding, this state vector can be described as a concatenation of $K$ row vectors (each has the dimension of $K$) of matrix $\mathbf{S}^M$. Each row $i$ of matrix $\mathbf{S}^M$ is associated with a user, whose $j$th element is
\begin{align}
\label{state_def}
\mathbf{S}^M_{ij} = 
\begin{cases}
\epsilon_j  \text{, if user } i \text{ has not cached file } f_j \\
0  \text{, if user } i \text{ has cached file } f_j \text{ and } i \ne j \\
0, \forall j=1,\ldots,K  \text{, if user } i \text{ has cached file } f_i
\end{cases}
.
\end{align}
where $\epsilon_j$ is the SINR threshold of the requested file of user $j$ (file $f_j$). If user $i$ has cached file $f_j$, the $j$th element on row $i$ will be 0 implying that user $i$ does not need to decode file $f_j$ in the superposed signal. If user $i$ has cached his requested file (file $f_i$), then all elements on row $i$ are 0 implying that user $i$ does not need to decode any file in the superposed signal. Once users generate their requests, the vector $\mathbf{S}$ is built at the BS, then based on the central controller decision, the BS gives back an action as a $K$-dimensional vector
\begin{align}
\mathbf{\alpha} = \left[\alpha_1 ~~ \alpha_2 ~~ \ldots ~~ \alpha_K \right]
\end{align}
which are the portions of power allocated to users' signal. Once power is allocated to signals sent to users, the number of users who successfully decode their desired signals is gathered by the BS as a reward which we aim to maximize in the long run.

\subsubsection{Deep-Reinforcement-Learning-Based Approach}

In the proposed learning algorithm hereafter, there are generally three phases called \textit{exploration}, \textit{training} and \textit{exploitation} phases. The description is given below.

\textbf{Exploration Phase}: In this phase, we try to discover which action returns the best reward for each encountered state. This is done by trying and observing. When an action (power allocation vector) is applied to the environment, the reward (number of success users) can be obtained correspondingly. Since the channel gains are random, the reward associated with a specific action in a specific state will also be random. Therefore, the average values of rewards will be considered instead of instantaneous values. This requires to apply a specific action to a specific state several times to obtain the average reward associated with that state-action pair. In this context, we are dealing with a continuous action space, hence, by generating actions randomly, there is no chance to pick an action twice. Even with quantization, it will take a significantly longer time. To address this problem, we denote $A_{\max}$ to be the maximum number of actions that we will apply to a given state before concluding about the best action associated with that state. The larger $A_{\max}$ is, the more reliable our conclusion about the best action is. 

Before executing the exploration phase, the users' channels are probed to construct the distribution of channels. Then, this phase is completed in four main steps: randomly generating users' requests and cached items, randomly generating a power allocation vector, drawing channel coefficients from the constructed distribution and computing the reward, and finally, computing the average reward and storing the best found action with its associated state. This process can be applied to a general context when the channel distribution is unknown. However, in this work we assume that the channel follows a Rayleigh distribution, hence, channel probing can be omitted. The mentioned four steps will be looped sequentially ($T_{\mathrm{Trial}}$ loops). Because the channels vary over time, the reward associated with a state-action pair is not consistent, hence, the third step (drawing channel coefficients and computing rewards) needs to be done several times to compute the average reward. Thus, it is put into an internal loop ($T_{\mathrm{Eval}}$ loops). The complexity of this process is, therefore, $O\left(T_{\mathrm{Trial}} \times T_{\mathrm{Eval}}\right)$. Note that we aim to deal with all possible cases, hence, we do not need to consider users' preferences or caching strategies, which is the reason why the first step in the loop can be done in a random manner. Algorithms \ref{explore} and \ref{explore_subfunc} summarize the described process in which $C_i^{\max}$ denotes the maximum cache capacity of user $i$ and $\alpha\left(\mathbf{S} \right)$ and $r\left(\mathbf{S} \right)$ are the best action and reward associated with state $\mathbf{S}$, respectively, that we have explored so far.

\begin{algorithm}
	\caption{\textbf{Exploration Phase}}\label{explore}
	\begin{algorithmic}[1]
	\State \textbf{Input}: $C_i^{\max} ~\forall i = 1, \ldots, K$ and $T_{\mathrm{Trial}}$.
	\State \textbf{Output}: A stored list of state-action pairs where the action in a pair is the best one found for the state in the same pair.
        \For{$t$ from 1 to $T_{\mathrm{Trial}}$}
        \State \textbf{Caching}: $K$ groups of files are chosen randomly. Group $i$ consists of an arbitrary number of different files up to $C_i^{\max}$ representing files cached by a user $i$.
        \State \textbf{Requesting}: $K$ files are chosen randomly representing files requested by users.
        \State \textbf{State Forming}: Based on the set of cached and requested files, the matrix of (\ref{state_def}) is constructed, and the state vector $\mathbf{S}$ is formed by concatenating row vectors.
        \State $\alpha^*, r^* \leftarrow$ \textit{find\_best\_action}$\left(\mathbf{S}\right)$
        \If{$\mathbf{S}$ has not been stored \textbf{and} memory is not full}
        \State $\mathbf{S}$ is stored; $\alpha\left(\mathbf{S} \right) \leftarrow \alpha^*$; $r\left(\mathbf{S} \right) \leftarrow r^*$
        \EndIf
        
        \If{$\mathbf{S}$ has been stored \textbf{and} $\alpha^* > \alpha\left(\mathbf{S} \right)$}
        \State $\alpha\left(\mathbf{S} \right) \leftarrow \alpha^*$; $r\left(\mathbf{S} \right) \leftarrow r^*$
        \EndIf
        \EndFor
	\end{algorithmic}
\end{algorithm}

\textbf{Training Phase}: After the exploration phase, we obtain a list of states and the corresponding best actions. Subsequently, a parametric function is needed to be designed to learn from this data set the general rule of allocating power. After training, the function will be used as a predictor to predict the best action for an input state. To support the design of the predictor, understanding properties of the training data is necessary. Particularly, because the list of good actions in the training data is constructed from the random acting process, it bears a certain randomness. The randomness of training data comes from the fact that there are many different action vector $\alpha$ resulting in the same reward for each state. Therefore, two very similar states can be associated with totally different action vectors, which causes inconsistency and obstructs the learning process.

For ease of understanding, an example is given as follows. Considering a three-user scenario, for a state $\mathbf{S}_1$, an action $\alpha_1=\left[0.7 ~~ 0.2 ~~ 0.1 \right]$ results in a failure of only user 3, which brings the best reward of $r\left(\mathbf{S}_1\right)=2$. However, another action $\alpha_2=\left[0.0 ~~ 0.3 ~~ 0.7 \right]$ which results in a failure of only user 1 can also give the best reward of $r\left(\mathbf{S}_1\right)=2$. Then, either $\alpha_1$ or $\alpha_2$ will be associated with $\mathbf{S}_1$ depending on which one is encountered first. In addition, there is a situation that a state $\mathbf{S}_2$ which is very similar to $\mathbf{S}_1$ goes with $\alpha_2$, while $\mathbf{S}_1$ goes with $\alpha_1$. Hence, for two very similar states, the actions going with them can be randomly different. Moreover, if channel conditions of users are good, they can decode the desired signals without being allocated much power. This means that action vectors such as $\alpha_3=\left[0.1 ~~ 0.15 ~~ 0.75 \right]$, $\alpha_4=\left[0.4 ~~ 0.51 ~~ 0.09 \right]$, and $\alpha_5=\left[0.9 ~~ 0.05 ~~ 0.05 \right]$ can all give the same best reward, although these vectors are very different and distributed randomly in the action space.

\begin{algorithm}
	\caption{\textit{find\_best\_action}$\left(\mathbf{S}\right)$}\label{explore_subfunc}
	\begin{algorithmic}[1]
		\State \textbf{Input}: A state $\mathbf{S}$, $A_{\max}$ and $T_{\mathrm{Eval}}$.
		\State \textbf{Output}: The best action for the input state and the corresponding reward. 
		\For{$j$ from 1 to $A_{\max}$}
		\State \textbf{Action Generating}: Generate $\alpha_j = \left(\alpha_{j1}, \ldots, \alpha_{jK} \right)$ randomly where $\alpha_{jk} \sim Unif\left(0, 1\right)$, $\forall k=1,\ldots,K$. If the $\left[K\left(k-1 \right) + k\right]$-th element of  $\mathbf{S}$ is greater than 0 then $\alpha_{jk} \leftarrow 0$. Finally, $\alpha_{jk} \leftarrow \alpha_{jk}/\sum_{i=1}^K \alpha_{ji} $.
		\For{$t$ from 1 to $T_{\mathrm{Eval}}$}
		\State \textbf{Reward Obtaining}: The channel coefficients are drawn from the obtained distribution (Rayleigh, in this case). From vectors $\mathbf{S}$ and $\alpha_j$, $r_{jt}$ is computed as the total number of success users following SIC process.
		\EndFor
		\State Estimate the average reward by: $\bar{r}_j = \frac{1}{T_{\mathrm{Eval}}}\sum_{t=1}^{T_{\mathrm{Eval}}}r_{jt}$.
		\EndFor
		\State $j^* \leftarrow \underset{j}{\text{argmax}} ~\bar{r}_j$
		\State \textbf{return} $\alpha_{j^*},\bar{r}_{j^*}$
	\end{algorithmic}
\end{algorithm}

Due to the complexity and noisiness of the training data in our work, and to generalize the relationship between states and actions, deep neural network models are built for the goal of anticipation. Our networks are trained based on the forward and backpropagation scheme \cite{BP:95}. The error is computed in the forward propagation stage, and the backpropagation stage evaluates the gradient of the error function with respect to the weight set via the derivative chain rule. By executing the forward and backpropagation alternatively, the weights can be updated in each training epoch. In the experiments in Section \ref{Sec:Results}, our networks are trained with two different types of loss. Also, the ADAM optimization algorithm \cite{Adam} is employed to minimize the loss. The first loss function type is simply the mean absolute error (MAE) between the network outputs and the target output vectors. The second type consists of two terms. The first term is simply the MAE between network outputs and target outputs. The second term is the average SINR values computed from the network outputs and target outputs. The average-SINR part consists of $K + \left(K-1\right) + \ldots + 1$ terms corresponding to SINR values associated with signals that users need to decode. For example, if $K=2$ with $\alpha_1 > \alpha_2$, the average SINR in this case will consist of three terms. The first one is the average SINR regarding signal 1 received by user 1, i.e., $\mathbb{E}\left[\frac{\left|h_1\right|^2\alpha_1P_{\max}}{\left|h_1\right|^2\alpha_2P_{\max}+\beta_1}\right]$. Since user 2 will need to decode signal 1 and then signal 2, the last two terms are $\mathbb{E}\left[\frac{\left|h_2\right|^2\alpha_1P_{\max}}{\left|h_2\right|^2\alpha_2P_{\max} + \beta_2}\right]$ and $\mathbb{E}\left[\frac{\left|h_2\right|^2\alpha_2P_{\max}}{\beta_2}\right]$.


Besides, we observe that the randomness takes place in two aspects of the training data (action vectors), the first one is the values of elements and the second one is the order of those elements. In order to mitigate the randomness and simplify the learning process, one of our solution is separating those two aspects and learning them separately. To be more specific, we design a dual-network model consisting of two deep networks with the same input and output dimensions, called $DNN_{val}$ and $DNN_{ord}$. We denote  $\alpha^{\mathbf{S}}_{val}$ and $\alpha^{\mathbf{S}}_{ord}$ the output from $DNN_{val}$ and $DNN_{ord}$, respectively, with input state $\mathbf{S}$. The elements of $\alpha^{\mathbf{S}}_{val}$ is supposed to indicate the power allocation for all users in a descending (or ascending) order. The elements of $\alpha^{\mathbf{S}}_{ord}$ is supposed to indicate the corrected order of elements in $\alpha^{\mathbf{S}}_{val}$. Note that elements of $\alpha^{\mathbf{S}}_{ord}$ have continuous values in $\left[0,1\right]$, and their order (not the values) are used to arrange elements in $\alpha^{\mathbf{S}}_{val}$ before applying to the environment.

\begin{figure*}[ht]
	\centering
	\includegraphics[scale = 0.31]{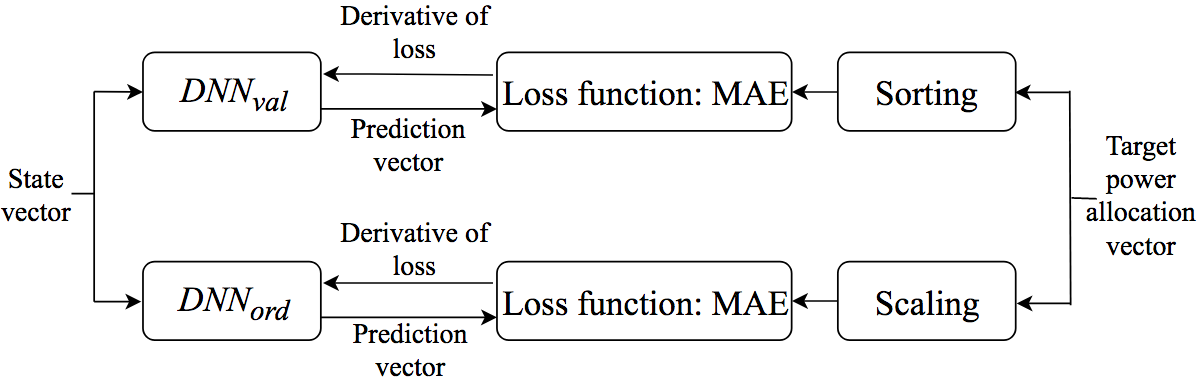}
	\caption{Illustration for the training phase of the dual-network prediction model. The network weights are optimized based on forward and backpropagation mechanism with MAE as a loss function.}
	\label{fig:DualNet_Training}
\end{figure*}

For training $DNN_{val}$, the set of target vectors are sorted to eliminate the randomness in element order. In our experiments in Section \ref{Sec:Results}, descending order is used. Then, $DNN_{val}$ is trained with the original input set and the sorted target output set. In terms of $DNN_{ord}$, we want to learn only the order of elements (to remove the randomness in their values). Hence, for the ease of learning and, concurrently, for preserving the correlation of data, we scale elements in the target output vectors by multiplying with $\xi_{scale} > 1$. The training process for the dual-network model is illustrated in Fig.~\ref{fig:DualNet_Training}.

\begin{figure*}[ht]
	\centering
	\includegraphics[scale = 0.30]{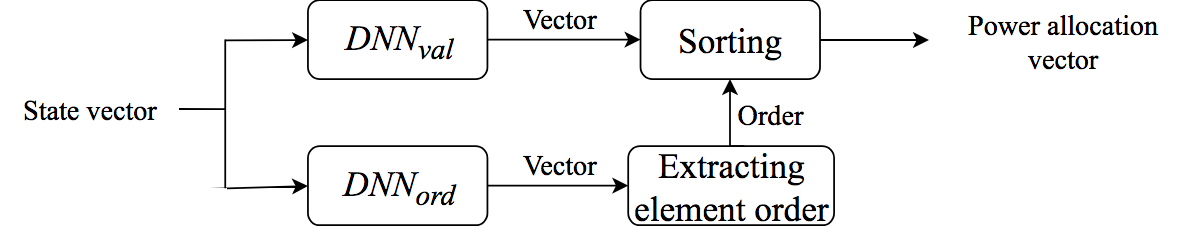}
	\caption{Exploiting dual-network model for predicting power allocation vector. }
	\label{fig:DualNet}
\end{figure*}

\textbf{Exploitation Phase}: Finally, the trained model can be used to perform the power allocation for every encountered state. The use of the single-network model is straightforwardly passing the input through layers of the network. Meanwhile, exploiting the dual-network model is a bit more complicated, thus, the process is summarized in Fig.~\ref{fig:DualNet}. Note that the order used in ``Sorting" and ``Extracting element order" must be the same as that used in the training phase.

In conclusion, from the timing perspective, solving (\ref{obj_temp})-(\ref{sec2_con6_formu1}) includes searching for the set of $\psi_{ij}$, whose delay is unstable with a high worst-case delay of $2^{K^2}T_{\alpha}$ where $T_{\alpha}$ is the delay of finding $\alpha_i$ for a given $\psi_{ij}$ set. For example, given the set $\psi_{ij}$, minimizing the Lagrangian function using a standard Newton's method will result in a complexity of $O\left( \varepsilon^{-2} \right)$ with $\varepsilon$ is the error between the obtained result and the nearest local optimum. In addition, the delay will take place infinitely many times whenever users make their requests. On the other hand, the proposed learning approach has a stable and finite one-time delay, because the delay comes mainly from the exploration and training phases. To be more specific, the complexity of the exploration phase is $O\left(T_{\mathrm{Trial}}\times T_{\mathrm{Eval}} \right)$ as mentioned, and the complexity of the training phase, with forward and backpropagation algorithm, is $O\left( cS_{train}T_{train} \right)$. Here, $c$ is a constant which depends on the number of layers and units of DNN model, $S_{train}$ is the training set size and $T_{train}$ is the number of training epochs. Finally, the complexity of the exploitation phase is $O\left(1 \right)$.

\section{Numerical Results}\label{Sec:Results}

Numerical results illustrating the performance of proposed methods are presented in this section. We consider a library of 38 files. Their SINR thresholds relative to the full bandwidth usage take values from $0.016$ to $0.608$ with a step of $0.016$, respectively. The default values of other parameters are as follows: $P_{\max}=1$, $\sigma^2=1$ (when using the entire bandwidth), $d_i = 1, \forall i=1,\ldots,K$ and every file has the same chance to be requested. Because the proposed methods are targeted to work for all cases of cache placement at the user side, in our experiments, the cache placement will be done in a random manner.

For the neural network architecture, we use a 5-layer network having the dimension of $9 \times 20$ (input layer), $20 \times 30$, $30 \times 20$, $20 \times 10$ (three hidden layers) and $10 \times 3$ (output layer), respectively. The activation functions between hidden layers are \textbf{relu} functions, and that at the output is a \textbf{softmax} function. For the case of the dual-network model, each network also has the same mentioned architecture. Those hyper parameters in our networks are obtained by tuning. The scaling factor when using the dual-network model is set to be $\xi_{scale}=2$ by default. At the beginning of the training session, the weights of our networks are initialized randomly.

\textbf{Baseline methods:} For performance evaluation, we include the following methods in our experiments:
\begin{itemize}
	\item Orthogonal multiple access (OMA): all users are allocated orthogonal subchannels and the power can be assigned following our Method 1 with only stage 1 (as in Fig.~\ref{fig:twostage_poweralloc}) where a user pair is replaced by an individual. Therefore, the expression (\ref{P_for_pairs}) can be applied with the following straightforward modifications: $\Psi^*_i = \lambda_i \beta_i \epsilon_i$ is associated with each user $i$ and the denominator is now $\sum_{j=1}^K \sqrt{\Psi_j^*}$.
	\item Equal power allocation: all users are allowed to share the whole bandwidth and allocated the same power of $P_{\max}/K$.
	\item Maximin-fairness power allocation (MMF) \cite{MMF}: MMF is chosen for comparison since both the network scenario and objective considered in this work are similar to ours. The differences are that there is no caching enabled, the objective is to maximize the lowest communication rate among users, and this method allocates power based on channel-to-noise ratios (CNRs) defined by $\frac{\left|h_i\right|^2}{d_i^\gamma \sigma^2}$. In other words, it requires the knowledge of instantaneous channel information which is not assumed to be available in our work. Therefore, in the experiments, the average CNRs will be used when applying this method.
\end{itemize}

\begin{figure}[ht]
	\centering
	\includegraphics[scale = 0.63]{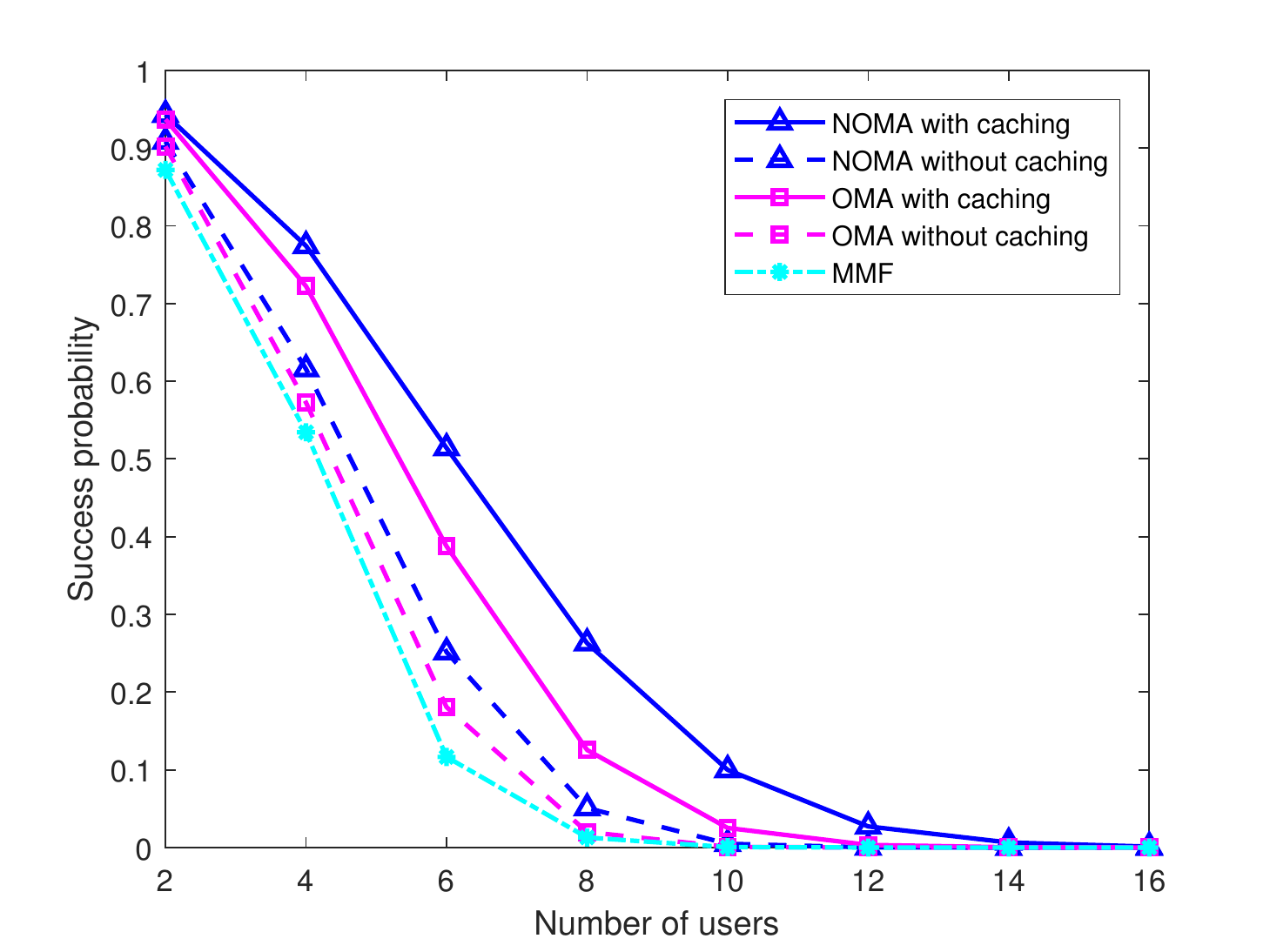}
	\caption{The success probability as a function of user number who joins the system under the deployment of Method 1. The comparison is conducted between NOMA and OMA with caching enabled and disabled. }
	\label{fig:M1_UserNumber}
\end{figure}

In Fig.~\ref{fig:M1_UserNumber}, we investigate the impact of the user density on the communication quality under the deployment of Method 1. There are $K=4$ users involved in this experiment. Users are paired with $\left|h_1\right|^2$ and $\left|h_2\right|^2$ follow the exponential distribution with mean 1 and 2, respectively. This setup is repeated for every pair. First of all, with caching enabled, the effectiveness of both NOMA and OMA is increased, it also enlarges the gap between these two schemes. This, once again, emphasizes the important role of caching in our communication systems. Secondly, as in the previous figure, NOMA still outperforms its counterpart. This is because each user can use as twice of the bandwidth as that of the OMA case, hence deal with lower SINR requirement. Although OMA offers a lower noise power for each user, it is not sufficient to compensate for the rise in SINR thresholds. In addition, the gap between NOMA and OMA is enlarged when caching in introduced, telling that NOMA scheme can make use better of the advantage of cache-enabled networks. Finally, the MMF method is a bit below the OMA scheme due to the unavailability of instantaneous channel information.

\begin{figure}
	\centering
	\subfigure[$10^5$-sample training set]{
		\includegraphics[scale = 0.63]{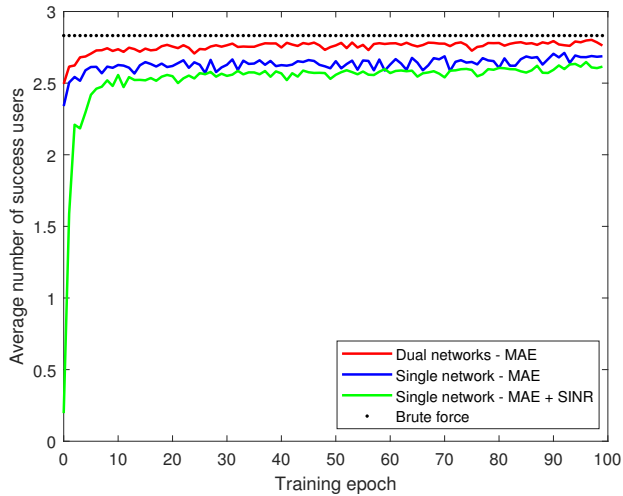}
	}
	~
	\subfigure[$10^4$-sample training set]{
		\includegraphics[scale = 0.63]{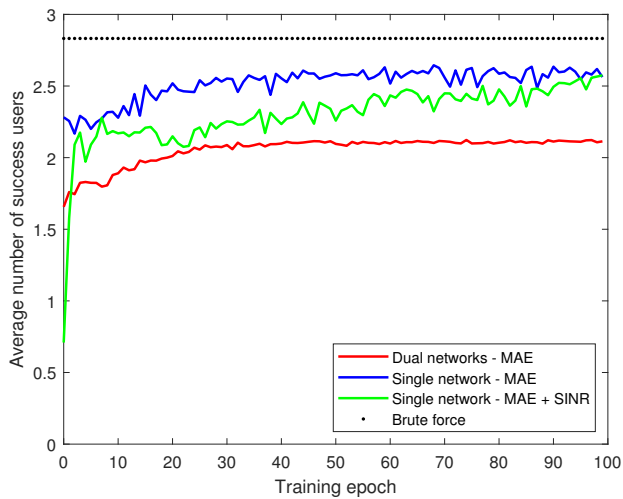}
	}
	\caption{Performance of prediction models in terms of the average number of success users during the training process.}
	\label{fig:M2_NetEpoch}
\end{figure}

The performance of Method 2 during the training session is recorded in Fig.~\ref{fig:M2_NetEpoch}. There are $K=3$ users in this context with $\left|h_1\right|^2, \left|h_2\right|^2$ and $\left|h_3\right|^2$ follow exponential distribution with mean 1, 2 and 3, respectively. Each user can cache up to 2 files. As mentioned previously, two kinds of predictor architectures are presented with different loss functions for training. ``MAE + SINR" implies the loss function with two terms of output's MAE and SINR's MAE. Note that in the system with 3 users, the SINR itself will consist of 6 terms. The average number of users who can successfully decode their desired signals is used as a metric to evaluate the prediction accuracy of neural networks. As can be seen from the figure, on one hand with a single network, we achieve slightly better results when using the simple MAE as loss function. This is because SINR depends on the order of power allocated amounts (stronger signals are decoded first and decoded by more users), hence, is a complicatedly non-differentiable function with respect to the network's output vector. On the other hand, using the dual networks gives the best results closing to brute force level (Fig.~\ref{fig:M2_NetEpoch}(a)). Since this model can reduce the uncertainty in the training data. However, the error from this kind of model is affected by that of two networks, thus, it is required to be trained well with sufficient samples. This makes the dual-network model appears to be the most sensitive one to dataset size. When the dataset size is restricted to one-tenth of the original one (Fig.~\ref{fig:M2_NetEpoch}(b)), the rise in the error of both networks suppresses the performance of this model. Generally, all the presented candidates have their accuracy grown when being trained with more samples.

\begin{figure}[ht]
	\centering
	\includegraphics[scale = 0.63]{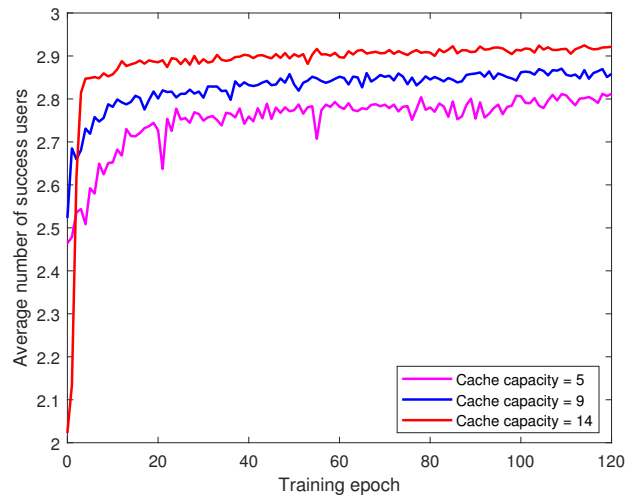}
	\caption{The performance of dual networks during the training process under different users' cache capacity conditions.}
	\label{fig:M2_CacheEpoch}
\end{figure}

Fig.~\ref{fig:M2_CacheEpoch} illustrates the performance of the dual-network models during its training process with different users' cache capacity conditions. The advantage of deploying caching at users' devices is not only associated with better results in terms of the average success user number, but also with the stability in the system performance as shown in the figure.

\begin{figure}[ht]
	\centering
	\includegraphics[scale = 0.63]{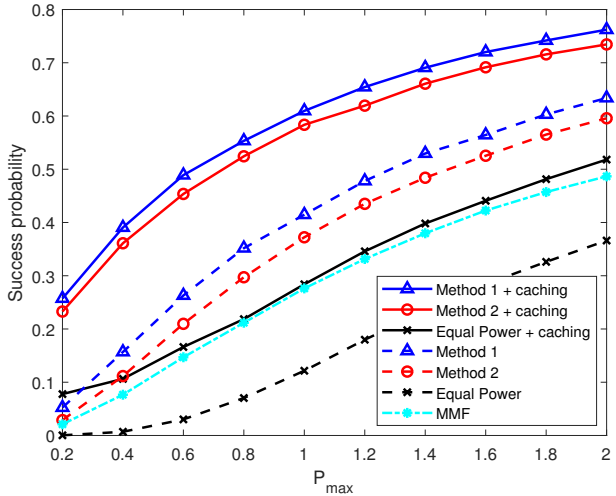}
	\caption{The adaptability of dual-network prediction model in a comparison to the optimal policy under the variation of communication SNR adjusted by the maximum transmission power $P_{\max}$.}
	\label{fig:Compare_Pmax}
\end{figure}

\begin{figure}[ht]
	\centering
	\includegraphics[scale = 0.63]{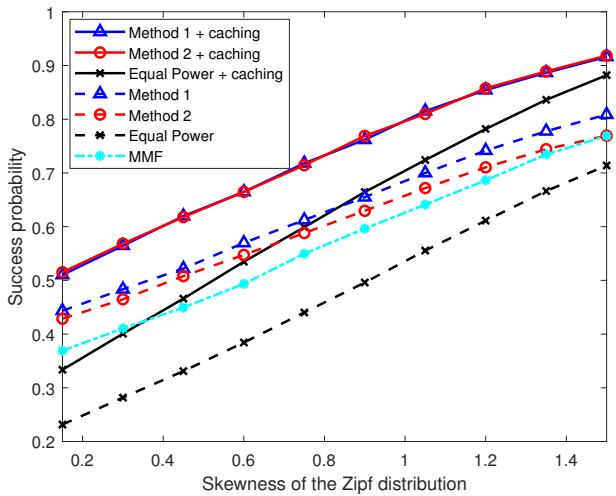}
	\caption{Under the assumption that the request probability follows Zipf distribution, this figure shows the adaptability of dual-network prediction model in comparison to the optimal policy with respect to the variation of the Zipf's skewness. When the skewness is large, users requests are focused on files with low indexes.}
	\label{fig:Compare_Zipf}
\end{figure}

In our experiments, the training stage of neural networks in Method 2 is done with respect to a specific communication condition of users. We would like to investigate how the trained model behaves when it is applied to a different context such as a different number of users and channel conditions. Regarding this, Fig.~\ref{fig:Compare_Pmax} and \ref{fig:Compare_Zipf} show the difference between the second method and the first method in a different environment than the one where neural networks are trained. Particularly, in these experiments, the number of users is 2 with $\left|h_1\right|^2$ and $\left|h_2\right|^2$ having their mean to be 1 and 2, respectively, with a variance of 1 and 4, respectively. Since the output from neural networks are fixed to have a dimension of 3, in order to work with 2-user case, we treat the third user as a virtual one who always caches his requests. This issue is a challenge for our prediction model resulted from the learning nature of parametric functions. When learning a general policy to maximize the reward associated with 3 users, the reward associated with the first two users will need to be sacrificed sometimes to achieve the best overall reward. Thus, training a new model to work directly with 2 users will give a better result. However, as mentioned we would like to examine the adaptability of our trained model in a new context. Another important point is that in this situation, the results from the first method is theoretically optimal, hence, is an upper bound for Method 2.

Fig.~\ref{fig:Compare_Pmax} presents the improvement in the success probability following the enhancement of SNR adjusted by rising $P_{\max}$. In Fig.~\ref{fig:Compare_Zipf}, the request probability of users is assumed to follow Zipf distribution with a certain skewness. This figure presents the effect from the skewness factor. As this factor increases, users' requests concentrate on the files with low indexes. Therefore, the optimal caching policy is to cache file from index 0 to 38 until the maximum capacity is reached. Although being brought to a different and unexpected situation, the prediction from our model is closed to that of the global optimal results, which verifies the effectiveness of our learning model. Furthermore, both of our methods are superior to all the baseline schemes. This is because the simple equal power allocation scheme does not exploit any of system information and the MMF method cannot adapt well to our context which has no instantaneous channel information.


\begin{figure}[ht]
	\centering
	\includegraphics[scale = 0.63]{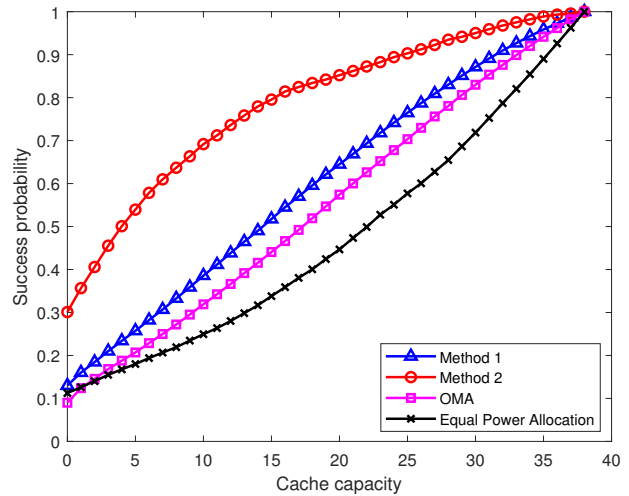}
	\caption{The performance comparison between Method 1, Method 2 and OMA scheme in a 4-user scenario. $\left|h_i\right|^2, \forall i=1,2,3,4$ follow the exponential distribution with mean 3 and variance 9.}
	\label{fig:Compare_4U}
\end{figure}

For a clearer comparison between methods, we train another dual-network prediction model for the case of 4 users. Each network of the predictor has 5 layers with dimensions of $16 \times 35$, $35 \times 50$, $50 \times 35$, $35 \times 12$ and $12 \times 4$. Other setups remain unchanged. The model is trained in a context where $\left|h_i\right|^2, \forall i=1,2,3,4$ follow the exponential distribution with mean 3 and variance 0.1, and the variance is 9 in a context of comparison. The corresponding results are in Fig.~\ref{fig:Compare_4U} which shows that regardless of the change in channel variance, Method 2 still outperforms the others with significant gaps. This is because this method exploits better the available bandwidth. However, Method 2 is more dependent on the system context than Method 1. Making use of orthogonal frequencies to isolate user pairs makes Method 1 simple and easy to be deployed in different contexts.

\section{Conclusion} \label{Sec:Conclusion}

In this work, we have combined caching and NOMA which are two prominent techniques in future wireless networks. Our analysis has shown that the combination creates another type of cache hit which takes place when users cache the requests of the others even without users' collaboration. This enables the interference cancellation, enhancing the effectiveness of both caching and NOMA. Such an interference-cancellation mechanism in a non-collaboration manner also helps simplify the system implementation, emphasizing that caching and NOMA should be deployed concurrently in future networks in order to boost the system performance further. Although deriving a joint cache placement and transmission strategy is not considered in this work, it is a topic of interest for further study.

In order to optimize the communication quality and concurrently guarantee the fairness among users, our target has been to maximize the probability that all users can successfully decode their desired signals. To achieve this, we have proposed divide-and-conquer-based and deep-learning-based methods. To compare the two methods, the former is simpler and more flexible to the system context with closed-form solutions derived. Although the second method requires system training, it is able to reach a higher performance due to better bandwidth usage. In this method, a dual DNN model has been proposed to overcome the noisiness/randomness problem in training data.

\section{Appendix: Proof of Theorem 1} \label{Sec:Appendix}
	$\blacksquare$ $1^{st}$ case: $\alpha \ge 0.5$.
	
	The conditions (\ref{conC1_1})-(\ref{conC1_3}) can be respectively rewritten as
	\begin{align}
	& \left|h_1\right|^2 \ge \frac{\epsilon_1\beta_1}{\alpha} \\
	& \left|h_2\right|^2\ge \max\left(\frac{\epsilon_1\beta_2}{\left(1+\epsilon_1\right)\alpha-\epsilon_1}, \frac{\epsilon_2\beta_2}{1-\alpha}\right) \\
	& ~ \alpha > 1 - \frac{1}{1+\epsilon_1}
	\end{align}
	
	 $\bullet$ $1^{st}$ subcase: $\frac{\epsilon_1\beta_2}{\left(1+\epsilon_1\right)\alpha-\epsilon_1} \ge \frac{\epsilon_2\beta_2}{1-\alpha}$. 
	
	The condition for this subcase is equivalent to
	\begin{align}
	\label{proof_conmax1}
	\alpha \le 1 - \frac{1}{1+\epsilon_1+\frac{\epsilon_1}{\epsilon_2}}.
	\end{align}
	
	Then, the system success probability is expressed as
	\begin{align}
	p^{C1}_{11} = \exp\left(-\frac{\lambda_1\epsilon_1\beta_1}{\alpha} - \frac{\lambda_2\epsilon_1\beta_2}{\left(1+\epsilon_1\right)\alpha-\epsilon_1}\right).
	\end{align}
	
	$p^{C1}_{11}$ is a monotonically increasing function of $\alpha$, hence, is maximized when
	\begin{align}
	\label{proof_result_1sub}
	\alpha = 1 - \frac{1}{1+\epsilon_1+\frac{\epsilon_1}{\epsilon_2}}.
	\end{align}
	
	$\bullet$ $2^{nd}$ subcase: $\frac{\epsilon_1\beta_2}{\left(1+\epsilon_1\right)\alpha-\epsilon_1} \le \frac{\epsilon_2\beta_2}{1-\alpha}$.
	
	The condition for this subcase is equivalent to
	\begin{align}
	\label{proof_conmax2}
	\alpha \ge 1 - \frac{1}{1+\epsilon_1+\frac{\epsilon_1}{\epsilon_2}}.
	\end{align}
	
	With the function $g_1^{\textit{C1}}$ defined in (\ref{for_appendix_1}), the success probability is
	\begin{align}
	p^{C1}_{12} = \exp\left(-g_1^{\textit{C1}}\left(\alpha\right)\right).
	\end{align}
	
	Maximizing $p^{C1}_{12}$ is finding $\alpha$ such that $g_1^{\textit{C1}}$ is minimized. Two stationary points can be obtained by solving $\frac{\partial g_2^{\textit{C1}}\left(\alpha\right)}{\partial \alpha}=0$, and it is straightforward to verify that the minimum is attained at the first point, $\alpha=1-\frac{1}{\sqrt{\zeta}+1}$. Combining the obtained result with (\ref{proof_conmax2}), we have
	\begin{align}
	\label{proof_result_2sub}
	\alpha = z_1^{\textit{C1}} = \max\left(1 - \frac{1}{\sqrt{\zeta}+1}, 1 - \frac{1}{1+\epsilon_1+\frac{\epsilon_1}{\epsilon_2}}\right).
	\end{align}
	Combining (\ref{proof_result_2sub}) with the result (\ref{proof_result_1sub}) from previous subcase yields the same result as (\ref{proof_result_2sub}), and this result satisfies $\alpha \ge 0.5$, since $\zeta \ge 1$.
	
	$\blacksquare$ $2^{nd}$ case: $\alpha \le 0.5$.
	
	The condition (\ref{conC1_1}) and (\ref{conC1_4}) can be rewritten as
	\begin{align}
	& \left|h_1\right|^2 \ge \frac{\epsilon_1\beta_1}{\alpha} \\
	& \left|h_2\right|^2 \ge \frac{\epsilon_2\beta_2}{1 - \left(1+\epsilon_2\right)\alpha} \\
	& \alpha < \frac{1}{1+\epsilon_2}
	\end{align}
	
	Then, with the function $g_2^{\textit{C1}}$ defined in (\ref{for_appendix_2}), the success probability is expressed as
	\begin{align}
	p^{C1}_{2} = \exp\left(-g_2^{\textit{C1}}\left(\alpha\right)\right).
	\end{align}
	
	Maximizing $p_{2}$ is equivalent to minimizing $g_2^{\textit{C1}}$. Similarly to the previous case, solving $\frac{\partial g_2^{\textit{C1}}\left(\alpha\right)}{\partial \alpha}=0$ and combing with $\alpha \le 0.5$, we have
	\begin{align}
	\label{proof_result_3}
	\alpha = z_2^{\textit{C1}} = \min\left(\frac{1}{1+\epsilon_2}\left(1 - \frac{1}{\sqrt{\zeta\left(1+\epsilon_2\right)}+1}\right), 0.5\right).
	\end{align} 
	
	To this end, to choose the better result between (\ref{proof_result_2sub}) and (\ref{proof_result_3}), we plug them into the corresponding success probability expression to compare. Note that in the case $\alpha \ge 0.5$ there are two subcases. However, at the optimal point (\ref{proof_result_1sub}) of the first subcase, we have $\frac{\epsilon_1\beta_2}{\left(1+\epsilon_1\right)\alpha-\epsilon_1} = \frac{\epsilon_2\beta_2}{1-\alpha}$. Therefore, it is sufficient to compare $g_1^{\textit{C1}}\left(z_1^{\textit{C1}}\right)$ and $g_2^{\textit{C1}}\left(z_2^{\textit{C1}}\right)$ as presented in Theorem \ref{theorem1}. 
	

\begin{thebibliography}{10}
	\providecommand{\url}[1]{#1}
	\csname url@samestyle\endcsname
	\providecommand{\newblock}{\relax}
	\providecommand{\bibinfo}[2]{#2}
	\providecommand{\BIBentrySTDinterwordspacing}{\spaceskip=0pt\relax}
	\providecommand{\BIBentryALTinterwordstretchfactor}{4}
	\providecommand{\BIBentryALTinterwordspacing}{\spaceskip=\fontdimen2\font plus
		\BIBentryALTinterwordstretchfactor\fontdimen3\font minus
		\fontdimen4\font\relax}
	\providecommand{\BIBforeignlanguage}[2]{{%
			\expandafter\ifx\csname l@#1\endcsname\relax
			\typeout{** WARNING: IEEEtran.bst: No hyphenation pattern has been}%
			\typeout{** loaded for the language `#1'. Using the pattern for}%
			\typeout{** the default language instead.}%
			\else
			\language=\csname l@#1\endcsname
			\fi
			#2}}
	\providecommand{\BIBdecl}{\relax}
	\BIBdecl
	
	\bibitem{vaezi2018multiple}
	M.~Vaezi, Z.~Ding, and H.~V. Poor, \emph{Multiple Access Techniques for {5G}
		Wireless Networks and Beyond}.\hskip 1em plus 0.5em minus 0.4em\relax
	Springer 2019.
	
	\bibitem{CAG:08:COMMAG}
	V.~Chandrasekhar, J.~Andrews, and A.~Gatherer, ``Femtocell networks: A
	survey,'' \emph{IEEE Commun. Mag.}, vol.~46, no.~9, pp. 59--67, Sep. 2008.
	
	\bibitem{DNSQ:18}
	K.~N. Doan, T.~V. Nguyen, T.~Q.~S. Quek, and H.~Shin, ``Content-aware proactive
	caching for backhaul offloading in cellular network,'' \emph{IEEE Trans.
		Wireless Commun.}, vol.~17, no.~5, pp. 3128 -- 3140, May 2018.
	
	\bibitem{YCTY:16:TWC}
	Y.~Shen, C.~Jiang, T.~Q.~S. Quek, and Y.~Ren, ``Device-to-device-assisted
	communications in cellular networks: An energy efficient approach in downlink
	video sharing scenario,'' \emph{IEEE Trans. Wireless Commun.}, vol.~15,
	no.~2, pp. 1575--1587, Feb. 2016.
	
	\bibitem{MGA_1}
	M.~Ji, G.~Caire, and A.~F. Molisch, ``Fundamental limits of caching in wireless
	{D2D} networks,'' \emph{IEEE Trans. Inf. Theory}, vol.~62, no.~2, pp.
	849--869, Feb 2016.
	
	\bibitem{DNSQ:Access}
	K.~N. {Doan}, T.~V. {Nguyen}, H.~{Shin}, and T.~Q.~S. {Quek}, ``Socially-aware
	caching in wireless networks with random {D2D} communications,'' \emph{IEEE
		Access}, vol.~7, pp. 58\,394--58\,406, May 2019.
	
	\bibitem{MWV:16:MAG}
	W.~Shin, M.~Vaezi, B.~Lee, D.~J. Love, J.~Lee, and H.~V. Poor, ``Non-orthogonal
	multiple access in multi-cell networks: Theory, performance and practical
	challenges,'' \emph{IEEE Commun. Mag.}, vol.~55, no.~10, pp. 176--183, Aug.
	2017.
	
	\bibitem{DBLP:journals/corr/WeiYNED16}
	Z.~Ding, Z.~Wei, J.~Yuan, D.~W.~K. Ng, and M.~Elkashlan, ``A survey of downlink
	non-orthogonal multiple access for {5G} wireless communication networks,''
	\emph{IEEE J. Sel. Areas Commun.}, vol.~35, no.~10, pp. 2181--2195, Oct.
	2017.
	
	\bibitem{ZMH:15:CL}
	Z.~Ding, M.~Peng, and H.~V. Poor, ``Cooperative non-orthogonal multiple access
	in {5G} systems,'' \emph{IEEE Commun. Lett.}, vol.~19, no.~8, pp. 1462--1465,
	Aug. 2015.
	
	\bibitem{YYATAK:13:VTC}
	Y.~Saito, Y.~Kishiyama, A.~Benjebbour, T.~Nakamura, A.~Li, and K.~Higuchi,
	``Non-orthogonal multiple access ({NOMA}) for cellular future radio access,''
	in \emph{Proc. IEEE Veh. Tech. Conf.}, Dresden, Germany, Jun. 2013.
	
	\bibitem{FHJV:16:TC}
	F.~Fang, H.~Zhang, J.~Cheng, and V.~Leung, ``Energy-efficient resource
	allocation for downlink non-orthogonal multiple access network,'' \emph{IEEE
		Trans. Commun.}, vol.~64, no.~9, pp. 3722 -- 3732, Sep. 2016.
	
	\bibitem{ZLJPX_4}
	Z.~Xiao, L.~Zhu, J.~Choi, P.~Xia, and X.~Xia, ``Joint power allocation and
	beamforming for non-orthogonal multiple access {(NOMA)} in {5G} millimeter
	wave communications,'' \emph{IEEE Trans. Wireless Commun.}, vol.~17, no.~5,
	pp. 2961--2974, May 2018.
	
	\bibitem{DSVPQ:18}
	K.~N. Doan, W.~Shin, M.~Vaezi, H.~V. Poor, and T.~Q.~S. Quek, ``Optimal power
	allocation in cache-aided non-orthogonal multiple access systems,'' in
	\emph{Proc. IEEE Int. Conf. Commun.}, Kansas City, MO, USA, May 2018, pp.
	1--6.
	
	\bibitem{DBLP:journals/corr/ZhuWHHYY17}
	J.~Zhu, J.~Wang, Y.~Huang, S.~He, X.~You, and L.~Yang, ``On optimal power
	allocation for downlink non-orthogonal multiple access systems,'' \emph{IEEE
		J. Sel. Areas Commun.}, vol.~35, no.~12, pp. 2744 -- 2757, Dec. 2017.
	
	\bibitem{DBLP:journals/corr/abs-1709-06951}
	Z.~Ding, P.~Fan, G.~K. Karagiannidis, R.~Schober, and H.~V. Poor, ``{NOMA}
	assisted wireless caching: Strategies and performance analysis,'' \emph{IEEE
		Trans. Commun.}, pp. 1--1, 2018.
	
	\bibitem{YFu18}
	Y.~{Fu}, H.~{Wang}, and C.~W. {Sung}, ``Optimal power allocation for the
	downlink of cache-aided noma systems,'' in \emph{Int. Conf. Wireless Commun.
		and Signal Process.}, Oct 2018, pp. 1--6.
	
	\bibitem{YFu19}
	Y.~{Fu}, Y.~{Liu}, H.~{Wang}, Z.~{Shi}, and Y.~{Liu}, ``Mode selection between
	index coding and superposition coding in cache-based noma networks,''
	\emph{IEEE Communications Letters}, vol.~23, no.~3, pp. 478--481, March 2019.
	
	\bibitem{CachingNOMA_2}
	Z.~Zhao, M.~Xu, W.~Xie, Y.~Li, and M.~Peng, ``A non-orthogonal multiple
	access-based multicast scheme in wireless content caching networks,''
	\emph{IEEE J. Sel. Areas Commun.}, vol.~35, no.~12, pp. 2723--2735, July
	2017.
	
	\bibitem{CachingNOMA_3}
	H.~Zhang, Y.~Qiu, K.~Long, G.~K. Karagiannidis, X.~Wang, and A.~Nallanathan,
	``Resource allocation in {NOMA}-based fog radio access networks,'' \emph{IEEE
		Wireless Commun.}, vol.~25, no.~3, pp. 110--115, July 2018.
	
	\bibitem{vaezi2019non}
	M.~Vaezi, R.~Schober, Z.~Ding, and H.~V. Poor, ``Non-orthogonal multiple
	access: Common myths and critical questions,'' \emph{IEEE Wireless Commun.},
	2019, to appear.
	
	\bibitem{MI_Prog_6}
	S.~Burer and A.~N. Letchford, ``Non-convex mixed-integer nonlinear programming:
	A survey,'' \emph{Surveys in Operations Research and Management Science},
	vol.~17, pp. 97--106, 03 2012.
	
	\bibitem{bookMIP_5}
	P.~Belotti, C.~Kirches, S.~Leyffer, J.~Linderoth, J.~Luedtke, and Ashutosh,
	``Mixed-integer nonlinear optimization,'' \emph{Acta Numerica}, vol.~22, pp.
	1--131, May 2013.
	
	\bibitem{bookMIP_7}
	J.~Lee and S.~Leyffer, ``Mixed-integer nonlinear programming,'' vol. 154, 2012.
	
	\bibitem{BP:95}
	Y.~Chauvin and D.~E. Rumelhart, Eds., \emph{Backpropagation: Theory,
		Architectures, and Applications}.\hskip 1em plus 0.5em minus 0.4em\relax
	Hillsdale, NJ, USA: L. Erlbaum Associates Inc., 1995.
	
	\bibitem{Adam}
	D.~P. Kingma and J.~Ba, ``Adam: A method for stochastic optimization,'' 2014.
	
	\bibitem{MMF}
	J.~{Zhu}, J.~{Wang}, Y.~{Huang}, S.~{He}, X.~{You}, and L.~{Yang}, ``On optimal
	power allocation for downlink non-orthogonal multiple access systems,''
	\emph{IEEE J. Sel. Areas Commun.}, vol.~35, no.~12, pp. 2744--2757, Dec 2017.
	
\end{thebibliography}


\end{document}